\documentclass[fleqn,twoside]{article}
\usepackage{espcrc2}

\usepackage{amsmath,amsfonts,amssymb}

\usepackage{graphicx}
\usepackage[figuresright]{rotating}

\title{Testing CPT Invariance with Neutrinos}

\author{Tommy Ohlsson\address[KTH]{Division of Mathematical Physics,
        Department of Physics, Royal Institute of Technology (KTH) -- Stockholm
        Center for Physics, Astronomy, and Biotechnology (SCFAB), \\ 
        Roslagstullsbacken 11, 106 91 Stockholm, Sweden}%
        \thanks{E-mail: {\tt tommy@theophys.kth.se}}%
        \thanks{In collaboration with: S.M.~Bilenky, M.~Freund,
        M.~Lindner, and W.~Winter. Talk presented at the XXXIst
        International Conference on High Energy Physics (ICHEP 2002),
        Amsterdam, The Netherlands, July 24-31, 2002.}
        }
       
\begin{document}

\begin{abstract}
We investigate possible tests of CPT invariance on the level of event
rates at neutrino factories. We do not assume any specific model, but
phenomenological differences in the neutrino-antineutrino masses and
mixing angles in a Lorentz invariance preserving context, which
could be induced by physics beyond the Standard Model. We especially
focus on the muon neutrino and antineutrino disappearance channels in
order to obtain constraints on the neutrino-antineutrino mass and
mixing angle differences. In a typical neutrino factory setup
simulation, we find, for example, that $|m_3 -
\overline{m}_3| \lesssim 1.9 \cdot 10^{-4} \, \mathrm{eV}$ and
$|\theta_{23} -\bar \theta_{23}| \lesssim 2^\circ$.
\vspace{1pc}
\end{abstract}

\maketitle

\section{INTRODUCTION}

The theme of the current presentation is to discuss the
plausibility of testing possible CPT invariance violation in the neutrino
sector using future so-called neutrino factories.

The CPT theorem is one of the milestones of local quantum field
theory (QFT). Moreover, the standard model (SM) of elementary
particle physics is in very good agreement with all (at
least most) existing experimental data. The CPT theorem is valid for the SM.
Thus, CPT violation is connected with the search for physics beyond
the SM. So far, no CPT violation has been found.
However, neutrinos have been suggested as a source of CPT violation.

Prior papers on CPT violation with neutrinos include
Refs.~\cite{Coleman:1997xq,Coleman:1998ti,Barger:2000iv,Murayama:2000hm}.
Recently, neutrinos as a source of CPT violation have been
investigated by several authors
\cite{Banuls:2001zn,Barenboim:2001ac,Xing:2001ys,Skadhauge:2001kk,Bilenky:2001ka,Barenboim:2002rv,Bahcall:2002ia,Strumia:2002fw,Mocioiu:2002pz,Barenboim:2002hx}.
Some of the models of CPT violation with neutrinos (beyond the SM) are
the following:
\begin{itemize}
\item S.R.~Coleman and S.L.~Glashow \cite{Coleman:1997xq,Coleman:1998ti}:
A Lorentz and CPT-violating model is introduced in which the most
general CPT-violating interaction $u^\dagger b u$ is allowed,
where $b$ is a Hermitian matrix. This interaction implies that the
energies of the ultra-relativistic neutrinos with definite momentum $p$ are the
eigenvalues of the matrix:
$$
cp + \frac{m^2}{2p} + b,
$$
where $m^2 \equiv m m^\dagger$ is the Hermitian mass squared matrix
($m$ is a complex symmetric mass matrix) and $c$ is also a Hermitian
matrix, which describes velocity-mixing. Limiting the discussion to
neutrino oscillations with two flavors, the neutrino oscillation
transition probability formula becomes
\begin{eqnarray}
&& {\rm P}(\nu_\alpha \to \nu_{\alpha'}) = 1 - \sin^2 2\theta \nonumber\\
&\times& \sin^2 \left[
  \left( \frac{\Delta m^2}{4 E} + \frac{\Delta b}{2} + \frac{\Delta c
    E}{2} \right) L \right], \nonumber
\end{eqnarray}
where $\Delta m^2$, $\Delta b$, and $\Delta c$ are the differences
between the eigenvalues of the matrices $m^2$, $b$, and $c$, respectively.
Note that in order for the above formula to hold the mixing angles
that diagonalize the matrices $m^2$, $b$, and $c$ need all to be equal
to each other.
\newpage
\item V.D.~Barger {\it et al.} \cite{Barger:2000iv}:
The effective Lorentz and 
  CPT-violating interaction for neutrinos is: $\bar{\nu}_L^\alpha
  b_{\alpha\alpha'}^\mu \gamma_\mu \nu_L^{\alpha'}$, which implies that
\begin{eqnarray}
\Delta {\rm P}_{\alpha\alpha}^{\rm CPT} &\equiv& {\rm P}(\nu_\alpha \to
\nu_\alpha) - {\rm P}(\bar\nu_\alpha \to \bar\nu_\alpha) \nonumber\\
&=& - 2 \sin^2 2\theta \sin \left(
\frac{\Delta m^2 L}{2E} \right) \nonumber\\
&\times& \sin (\Delta b L) \nonumber
\end{eqnarray}
when the matrices $m^2$ and $b$ are diagonalized by the same mixing
angle $\theta$.
\item G.~Barenboim {\it et al.} \cite{Barenboim:2001ac}: CPT violation (but
no Lorentz violation) is suggested to be so
strong that the mass spectra of neutrinos and antineutrinos are
completely different. This means that it would be possible to describe solar,
atmospheric, and LSND neutrino data at the same time. Such a model is
accomplished by changing the Hamiltonian from
$$
H_0 = \int \frac{d^3 p}{(2\pi)^3} ({\bf p}^2 + m^2) \sum_s \left[
  {a_{\bf p}^s}^\dagger a_{\bf p}^s + {b_{\bf p}^s}^\dagger b_{\bf p}^s
\right]
$$
to
\begin{eqnarray}
H_0 &=& \int \frac{d^3 p}{(2 \pi)^3} \sum_s \Big[ ({\bf p}^2 + m^2)
  {a_{\bf p}^s}^\dagger a_{\bf p}^s \nonumber\\
&+& ({\bf p}^2 + {\overline{m}}^2)
  {b_{\bf p}^s}^\dagger b_{\bf p}^s \Big], \nonumber
\end{eqnarray}
where $m \neq \overline{m}$, {\it i.e.}, the mass of a neutrino is not
equal to the mass of the corresponding antineutrino. For $m \neq
\overline{m}$ the new Hamiltonian violates CPT invariance and
locality, since there is no possibility to derive it from any local QFT.
\end{itemize}

\section{CPT TESTS AT A NEUTRINO FACTORY}

As we have seen, CPT violation implies physics beyond local QFT, which
means that we have to consider Planck scale physics, large extra dimensions,
or string theory. In Ref.~\cite{Barenboim:2001ac}, CPT violation was used to
accommodate the LSND result
\cite{Athanassopoulos:1996jb,Athanassopoulos:1998pv,Aguilar:2001ty}.
Here we will study precision measurements
at a future neutrino factory, which will lead to interesting
phenomenological limits on the differences in the
neutrino-antineutrino masses and mixing angles.

Now, CPT invariance implies that
\begin{equation}
{\rm P}(\nu_\alpha\to\nu_{\alpha'})= {\rm
  P}(\bar\nu_{\alpha'}\to\bar\nu_{\alpha}),
\end{equation}
where ${\rm P}(\nu_\alpha \to \nu_{\alpha'})$ is the neutrino
oscillation transition probability that $\nu_\alpha \to \nu_{\alpha'}$
will occur. In the present discussion, we will violate CPT
invariance and locality (but not necessarily Lorentz invariance) by
assuming that masses and mixings differ for neutrinos and antineutrinos.
For neutrinos we have $m_a$ and $U$, whereas for antineutrinos we have
$\overline{m}_a$ and $\bar U$. Thus, as a consequence,
\begin{equation}
{\rm P}(\nu_\alpha\to\nu_{\alpha'}) \neq
{\rm P}(\bar\nu_{\alpha'}\to\bar\nu_{\alpha}),
\end{equation}
where for two neutrino flavors
\begin{eqnarray}
{\rm P}(\nu_\alpha \to \nu_{\alpha'}) &\equiv& \delta_{\alpha\alpha'}
\nonumber\\ 
&-& (2\delta_{\alpha\alpha'} - 1) \sin^2 2\theta \sin^2 \frac{\Delta m^2 L}{4
    E}, \nonumber\\
{\rm P}(\bar\nu_\alpha \to \bar\nu_{\alpha'}) &\equiv&
\delta_{\alpha\alpha'} \nonumber\\
&-& (2\delta_{\alpha\alpha'} - 1) \sin^2
2\bar\theta \sin^2 \frac{\Delta \overline{m}^2 L}{4 E}, \nonumber
\end{eqnarray}
which means that
\begin{eqnarray}
\Delta {\rm P}_{\alpha\alpha}^{\rm CPT} &\equiv& {\rm P}(\nu_\alpha \to
\nu_\alpha) - {\rm P}(\bar\nu_\alpha \to \bar\nu_\alpha) \nonumber\\
&=& - \sin^2 2\theta \sin^2
\frac{\Delta m^2 L}{4E} \nonumber\\
&-& \sin^2 2\bar\theta \sin^2 \frac{\Delta
  \overline{m}^2 L}{4E}. \nonumber
\end{eqnarray}
For three neutrino flavors we have more
complicated formulas. However, we will carry out our numerical calculations
using three neutrino flavors.

At a neutrino factory, neutrinos would be produced in muon decays
$\mu^+ \to e^+ \nu_e \bar\nu_\mu$ (or $\mu^- \to e^- \bar\nu_e \nu_\mu$).
The straightforward test of CPT violation would be to check the
appearance relation ${\rm P}(\nu_e \to \nu_\mu) = {\rm P}(\bar\nu_\mu
\to \bar\nu_e)$ [or ${\rm P}(\bar\nu_e \to \bar\nu_\mu) = {\rm P}(\nu_\mu
\to \nu_e)$].
However, this would require to measure the sign of the charge of the
produced lepton, which could be hard. We propose instead to check the equality
\begin{equation}
{\rm P}(\nu_\mu \to \nu_\mu) = {\rm P}(\bar\nu_\mu \to \bar\nu_\mu),
\end{equation}
{\it i.e.}, the $\nu_\mu$ and $\bar\nu_\mu$ disappearance channels.
The advantages of checking these disappearance channels are, for example:
\begin{itemize}
\item high event rates, \vspace{-2.5mm}
\item no beam contamination, {\it i.e.}, no relevant background is
  present, \vspace{-2.5mm}
\item small matter effects, \vspace{-2.5mm}
\item large neutrino oscillation effects.
\end{itemize}
Thus, we could obtain exclusion limits for tiny CPT-violating effects.

If we consider the $\nu_\mu$ and $\bar\nu_\mu$ channels as independent
experiments, then CPT violation in neutrino oscillations can be
quantified by the following asymmetries:
\begin{eqnarray}
\delta &\equiv& |\Delta m^2_{32} - \Delta \overline{m}^2_{32}|,\\
\epsilon &\equiv& | \sin^2 2 \theta_{23} - \sin^2 2 \bar\theta_{23}|.
\end{eqnarray}
For $m_1 \ll \sqrt{\Delta m^{2}_\odot}$ (hierarchical) and 
$|m_3 - \overline{m}_3|\ll {(m_{3})_{\rm average}}$, we have
\cite{Bilenky:2001ka}
\begin{eqnarray}
\delta &\simeq& 2 \, a_{\rm{CPT}} \, \Delta m^{2}_{32},\\
\epsilon &\simeq& 2 \, b_{\rm{CPT}} \, \sqrt{\sin^2 2\theta_{23}} \,
\sqrt{1 - \sin^2 2\theta_{23}} \nonumber\\
&\times& \arcsin \sqrt{\sin^2 2\theta_{23}},
\end{eqnarray}
where the asymmetry parameters are
$$
a_{\rm{CPT}} \equiv \frac{|m_3 - \overline{m}_{3}|}{(m_{3})_{\rm
    average}}
$$
and
$$
b_{\rm{CPT}} \equiv \frac{\left|\theta_{23} - \bar\theta_{23}\right|}
{(\theta_{23})_{\rm average}}.
$$

Next, we want to estimate the sensitivities $\delta a_{\rm CPT}$ and
$\delta b_{\rm CPT}$ of the asymmetry parameters $a_{\rm CPT}$ and
$b_{\rm CPT}$. The sensitivities to possible CPT violation are given
by the accuracies with which $a_{\rm CPT}$ and $b_{\rm CPT}$ can be
measured. Comparing $a_{\rm CPT}$ and $b_{\rm CPT}$ with the
corresponding relative statistical errors $\delta \Delta m_{32}^2$ and
$\delta \theta_{23}$ of the measurements of $\Delta m_{32}^2$ and
$\theta_{23}$, we obtain the sensitivities of the asymmetry parameters
\cite{Freund:2001ui}. Thus, the sensitivities $\delta a_{\rm CPT}$ and $\delta
b_{\rm CPT}$ for the asymmetry parameters $a_{\rm CPT}$ and $b_{\rm
  CPT}$ are given by:
\begin{eqnarray}
 \delta a_{\rm{CPT}} &\sim& \frac{\delta \Delta m_{32}^2}{2}, \\
 \delta b_{\rm{CPT}} &\sim& \delta \theta_{23}.
\end{eqnarray}

\begin{figure}[ht!]
\begin{center}
\vspace{5mm}
\includegraphics[width=5cm,angle=-90]{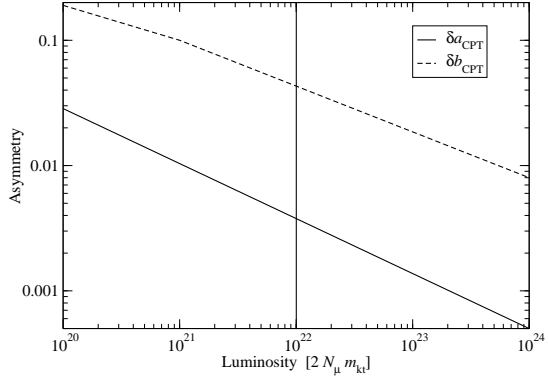} 
\vspace{-5mm}
\end{center}
\caption{The sensitivities $\delta a_{\rm CPT}$ (solid curve) and
$\delta b_{\rm CPT}$ (dashed curve) of an estimate of the asymmetries
$a_{\rm CPT}$ and $b_{\rm CPT}$ as a function of the luminosity ${\cal
L} \equiv 2 N_\mu m_{\rm kt}$. Parameter values: $E_\nu = 50 \,
\mathrm{GeV}$ (muon energy), $L = 3000 \, \mathrm{km}$ ($\Delta
m^2_{23}$, solid curve) / $L = 7000 \, \mathrm{km}$ ($\theta_{23}$,
dashed curve), $m_{\rm kt} = 10 \, \mathrm{kt}$ (mass of detector),
$10^{20}$ muons/year, and 5 years (running time). The figure has been
adopted from Ref.~\cite{Bilenky:2001ka}.}
\label{fig:1}
\end{figure}

The result of a possible future neutrino factory setup simulation is
presented in Fig.~\ref{fig:1}.
For our numerical calculations we assumed a 10 kt detector and
$10^{20}$ stored muons per year during 5 years.
Using the obtained upper bounds in Fig.~\ref{fig:1} ($a_{CPT}
\lesssim 3.8 \cdot 10^{-3}$ and $b_{CPT} \lesssim 4.3 \cdot 10^{-2}$)
as well as $(m_3)_{\rm average} \simeq \sqrt{\Delta m_{\rm atm}^2}
\leq 5 \cdot 10^{-2}~\mathrm{eV}$ and $(\theta_{23})_{\rm average}
  \simeq \theta_{\rm atm} = 45^\circ$, we find that
\begin{eqnarray}
|m_3 -\overline{m}_3| &\lesssim& 1.9 \cdot 10^{-4}~\mathrm{eV},\\
|\theta_{23} -\bar \theta_{23}| &\lesssim& 2^\circ,
\end{eqnarray}
which correspond to
\begin{eqnarray}
a_{\rm CPT} &\lesssim& 0.38 \%,\\
b_{\rm CPT} &\lesssim& 4.3 \%.
\end{eqnarray}

\section{SUMMARY \& CONCLUSIONS}

In summary, CPT violation is not allowed in local QFT. Thus,
fundamental CPT violation would mean physics beyond local QFT and the SM,
such as Planck scale physics, large extra dimensions, or string
theory. Finally, we have shown in our neutrino factory setup
simulation that CPT violation is detectable if
$|m_3 -\overline{m}_3| \not\lesssim 1.9 \cdot 10^{-4}~\mathrm{eV}$ and
$|\theta_{23} -\bar \theta_{23}| \not\lesssim 2^\circ$.

\section*{Acknowledgments}

I would like to thank my co-workers Samoil M. Bilenky, Martin Freund,
Manfred Lindner, and Walter Winter for fruitful collaboration and
H{\aa}kan Snellman for proof-reading this proceeding.

This work was supported by the Swedish Foundation for International
Cooperation in Research and Higher Education (STINT), the Wenner-Gren
Foundations, the Swedish Research Council (Vetenskapsr{\aa}det),
Contract No. 621-2001-1611, and the Magnus Bergvall Foundation
(Magn. Bergvalls Stiftelse).

\end{document}